\documentclass[aps,prd,preprint,showpacs,preprintnumbers,amsmath,amssymb, superscriptaddress]{revtex4-1}
\usepackage{graphicx} 
\usepackage{xcolor}
\usepackage{amsmath,amssymb,ulem}
 \usepackage{hyperref}
\usepackage[top = 2cm, bottom = 2cm, right = 2cm, left =2cm]{geometry}

\usepackage{hyperref}
\hypersetup{
    colorlinks,
    citecolor=blue,
    filecolor=green,
    linkcolor=purple,
    urlcolor=red,
}
\usepackage{slashed}
\newcommand{\meio}{{}^1\!/{}_{\!2}}
\newcommand{\tmeio}{{}^3\!/{}_{\!2}}

\usepackage{hyperref}
\hypersetup{colorlinks,breaklinks, citecolor=[rgb]{0,0.0,1.0}, urlcolor=[rgb]{0.0,0.0,1.0}, linkcolor=[rgb]{0,0.5,0.9}}

\def\rf{\eqref}

\def\lal{&&\ {}}

\def\beq{\begin{equation}}
\def\eeq{\end{equation}}
\def\bear{\begin{eqnarray}}
\def\bearr{\begin{eqnarray} \lal}
\def\ear{\end{eqnarray}}
\def\earn{\nonumber \end{eqnarray}}

\def\nnn{\nonumber\\ \lal }

\def\yyy{\\[5pt] \lal }

\def\diag{\,{\rm diag}\,}

\def\d{\partial}
\usepackage{hyperref}
\hypersetup{colorlinks,breaklinks,
			citecolor=[rgb]{0,0.0,1.0},
            urlcolor=[rgb]{0.0,0.0,1.0},
            linkcolor=[rgb]{0,0.5,0.9}}

\begin{document}

\title{Static and stationary loop quantum black bounces}


\author{C. R. Muniz}
\email{celio.muniz@uece.br}
\affiliation{Faculdade de Educação, Ciências e Letras de Iguatu, CEP 63.502-253, Iguatu-CE - Brazil}
\author{G. Alencar}
\email{geova@fisica.ufc.br}
\affiliation{Department of Physics, Universidade Federal do Cear\'a (UFC), Campus do Pici,  C.P. 6030, CEP 60455-760, Fortaleza-CE - Brazil}
\author{M. S. Cunha}
\email{marcony.cunha@uece.br}
\affiliation{Universidade Estadual do Cear\'a (UECE), Centro de Ciências e Tecnologia,  CEP 60.714.903, Fortaleza-CE - Brazil.}
\author{Gonzalo J. Olmo}
\email{gonzalo.olmo@uv.es}
\affiliation{Departamento de Física Teórica and IFIC, Centro Mixto Universitat de València–CSIC, Universitat de València, 46100 Burjassot, València, Spain}
\affiliation{Department of Physics, Universidade Federal do Cear\'a (UFC), Campus do Pici,  C.P. 6030, CEP 60455-760, Fortaleza-CE - Brazil}
\date{\today}

\begin{abstract}

We explore a static and stationary black bounce geometry inspired by Loop Quantum Gravity (LQG), focusing on how LQG corrections and regularization parameters affect its properties. Building on the spherically symmetric and static black hole solution from \cite{Kelly:2020uwj}, we trace its origin to Non-Linear Electrodynamics (NED) with electric and magnetic charges and check the energy conditions (NEC, WEC, SEC). By extending the geometry using the Simpson-Visser procedure, we construct a black hole-wormhole bounce structure, influenced by LQG parameters. We analyze the horizon structure to constrain parameters for black holes and wormholes, and examine curvature and new sources including a phantom-type scalar field to ensure spacetime regularity and adherence to energy conditions. Thermodynamic properties are also studied, revealing the existence of remnants and phase transitions. Additionally, we derive a rotating black bounce solution, verifying its regularity, and putting forward that the spherical bouncing surface turns into an ellipsoid with no ring singularity. Finally, we find that increasing the LQG parameter leads to smaller ergospheres and reduced shadows, with potential implications for observational astrophysics and quantum gravitational signatures.

\end{abstract}


\maketitle

\section{Introduction}
Recent advances in observational technologies have revitalized interest in gravitation theory. High-precision measurements have led to the direct detection of gravitational waves by the LIGO-VIRGO-KAGRA observatories,  to the imaging of supermassive black holes by the Event Horizon Telescope (EHT), and to precise determination of cosmological 
parameters, significantly enhancing our understanding of the universe and offering new challenges to our theories \cite{LIGOScientific:2016aoc, LIGOScientific:2017vwq, EventHorizonTelescope:2022wkp, EventHorizonTelescope:2019dse,  Planck:2018vyg, ACT:2020gnv, SDSS:2006srq}. Nonetheless, despite the  success in the observational determination of model parameters, there are fundamental reasons to believe that General Relativity (GR) may not account for all the possible gravitational effects. Consequently, the search for modified theories of gravity has attracted significant attention from the theoretical physics community. 

In line with this, new black hole solutions have recently been found by applying the quantization techniques of Loop Quantum Gravity to spherically symmetric solutions \cite{Kelly:2020uwj, Lewandowski:2022zce}, following the success of  Loop Quantum Cosmology models \cite{Ashtekar:2003hd,Ashtekar:2006rx,Ashtekar:2006wn,Ashtekar:2007em,Assanioussi:2019iye}. To this end, an effective Hamiltonian and a corresponding metric are introduced to describe the vacuum exterior solution within the context of the Oppenheimer-Snyder (OS) collapse model, specifically in spherically symmetric spacetimes \cite{Kelly:2020uwj}. A novel approach has also been proposed for the quantum OS model, which bridges the interior and exterior regions. This method incorporates LQC corrections within the interior spacetime, extending them to the exterior via specific boundary matching conditions \cite{Lewandowski:2022zce}.

In particular, in ref. \cite{Kelly:2020uwj} the authors found a line element of the form
\begin{equation}
ds^2 =f(r) dt^2 - f(r)^{-1} dr^2 - r^2 d\Omega^2, \;f(r) = 1 - \frac{2M}{r} + \frac{\alpha^2 M^2}{r^4}, \label{eq1}
\end{equation}
where the domain of the radial coordinate is bounded from below to impose a minimum radius able to cure the singularity. This minimal radius is given by
\begin{equation}
r>r_{b}= (\alpha^2 M/2)^{1/3},\label{bounce-alpha}
\end{equation}
where $\alpha^2=16\sqrt{3}\pi\gamma^3 \ell^2_P$, with $\ell_P$ being the Planck length and $\gamma$ is the Barbero-Immirzi parameter. The above cut-off is found in other models of LQG \cite{Gambini:2020nsf}. 
One can check that the line element (\ref{eq1}) has two horizons located approximately at 
\begin{equation}
r_{inner}= \left(\frac{\alpha^2 M}{2}\right)^{1/3} \!+ \ \frac{1}{6}\left(\frac{\alpha^4}{4M}\right)^{1/3}\!\!\!\!\!\!,\quad r_{outer}= 2M -\frac{\alpha^2}{8M}\label{horizons}.
\end{equation}
The properties of the above black hole have been studied, finding that the solution is stable under scalar and vector perturbations by computing the quasinormal modes, while the shadows were studied in \cite{Yang:2022btw}. This space-time has been  generalized to the LQG-AdS case, and its thermodynamics was also studied \cite{Wang:2024jtp}.

An undesired feature of the line element (\ref{eq1}) is that the domain of the radial coordinate $r$ can be naturally extended to $r\to 0$, leading to a singular space-time. Thus, an analytical way to avoid the singularity without artificially restricting the domain of $r$ would be very valuable. A natural way out of this problem is provided by the Simpson-Visser procedure \cite{Simpson:2018tsi}, which consists on the replacement \( r \to \sqrt{r^2 +a^2} \), where $a$ is a regularization parameter. If we choose $ a=r_b$, we get the desired regular solution with the same properties as the original one in central and far away regions. Furthermore, if we keep $a$ as a free parameter, we can get LQG-inspired wormholes and regular black holes by suitably tunning the value of $a$. It is in this extended scenario where we develop the main idea of this work: the Loop Quantum Black Bounce (LQBB). Our approach ensures the absence of the singularity and retains the main qualitative aspects of the LQG-inspired solution (\ref{eq1}). 

To better understand the inner workings of the Simpson-Visser procedure, it will be instructive to study the effective sources that generate the line element (\ref{eq1}) and of its transformed LQBB version. The thermodynamics of the LQBB model will also be studied and compared with the case of Schwarzschild. We will also consider the extension of the LQBB line element to the rotating case, paying special attention to the fate of the ring singularity that arises in the usual Kerr black hole. 


The paper is structured as follows: In Section II, we identify the source for the exterior solution of the LQG black hole. Section III discusses the application of Simpson-Visser regularization, where we derive the quantum loop black bounce solution and examine the horizons and energy conditions. In Section IV, we determine the sources for the regularized case. Section V delves into the thermodynamics of the solutions, while Section VI extends the analysis to the rotating case and investigates the resulting horizon structure, ergospheres, regularity, and shadows. Finally, Section VII presents our conclusions.

\section{LQG-inspired black hole} \label{sec_2}
As a warm up for what will come, we first revisit the solution of Ref. \cite{Kelly:2020uwj} by computing the effective energy sources that generate that line element,  studying the corresponding energy conditions, and determining their nature in terms of fundamental fields. 

\subsection{Energy Conditions}\label{EC}

Applying the coordinate-independent descriptions of the energy conditions — namely, the null energy condition (NEC), the weak energy condition (WEC), and the strong energy condition (SEC)—to the LQG-inspired black hole, we have
\begin{enumerate}
    \item NEC is defined as $T_{\mu \nu} k^{\mu} k^{\nu} > 0$ for all null vectors $k^{\mu}$ which provides the condition 
    \begin{eqnarray}
        \rho + p_r \geq 0.
    \end{eqnarray}
    \item WEC is defined as $T_{\mu \nu} t^{\mu} t^{\nu} > 0$ for all timelike vectors $t^{\mu}$ which provides the conditions
    \begin{eqnarray}
        \rho &\geq& 0, \\
        \rho + p_r &\geq& 0
    \end{eqnarray}
    \item SEC is defined as $\left( T_{\mu \nu} - \frac{1}{2} T g_{\mu \nu} \right) t^{\mu} t^{\nu} \geq 0$ which provides the conditions
    \begin{eqnarray}
        \rho + p_r+ 2p_t &\geq& 0. \\
      \end{eqnarray}
\end{enumerate}
Inserting the metric of Eq. (\ref{eq1}) into Einstein's equations and interpreting the result in terms of an effective fluid, we find 
\begin{equation}
  \rho=\frac{3 \alpha ^2 M^2}{r^{6}},\;p_r= -\frac{3 \alpha ^2 M^2}{r^{6}},\;p_t= \frac{6 \alpha ^2 M^2}{r^{6}}.
\end{equation}
From this result and the previous definitions, one can verify that all the energy conditions are satisfied. The above results show that we do not need exotic matter as a source. In the next section, we will find a source for the LQG black hole. 
%
\subsection{Source for the LQG-inspired black hole}
To identify the effective source for the LQG black hole, we first note that, for a general metric function $f$, the corresponding components of the Einstein tensor are:
            \bear \label{ET}
		G^0_0 &=& G^1_1 = -\frac{f'(r) }{r}-\frac{f(r) }{r^2}+\frac{1}{r^2},\\
		G^2_2 &=& G^3_3 =-\frac{f'(r) }{r}-\frac{f''(r)}{2}.
            \ear    
A general source that has the above algebraic structure is an electromagnetic field. However, the usual Maxwell action provides a dependence of the form $1/r^2$ in the metric, which is not what we need. It is thus necessary to consider nonlinear electrodynamics (NED) sources. Accordingly, we propose the total action in the form
\beq
    S=\int \sqrt{-g}d^4x\left[\frac{R}{2\kappa^2 } - L(F)\right]\ ,    			\label{Action}
\eeq
where $g$ is the determinant of the metric $g_{\mu\nu}$, $\kappa^2=8\pi G/c^4$ (with Newton's constant $G\to 1$ and the speed of light $c\to 1$ in our system of units), $L(F)$ is the NED Lagrangian, 
  $F=\frac{1}{4}F^{\mu\nu}F_{\mu\nu}$, and $F_{\mu\nu} = \d_\mu A_\nu - \d_\nu A_\mu$ is the 
  electromagnetic field tensor. 

The corresponding field equations are obtained by varying the action \eqref{Action} with respect  $A_\mu$, and $g^{\mu\nu}$, leading to
\bearr
    \nabla_\mu \left[L_F F^{\mu\nu}\right]=0,\quad \nabla_\mu^{*}  F^{\mu\nu}=0, 
\label{eq-F}
\\   \lal
    G_{\mu\nu}=R_{\mu\nu} - \frac 12 g_{\mu\nu} R = \kappa^2 T_{\mu\nu} ^{EM}, 
\label{eq-Ein}
\ear
where $L_F=d L/d F$, the asterisk denotes the Hodge dual,  and $T_{\mu\nu}^{EM}$ is the stress-energy tensor for the electromagnetic field,  which is given by:
\bearr                   \label{SET-F}
    T_{\mu\nu} ^{EM} =  g_{\mu\nu} L(F ) - L_F {F_{\nu}}^{\alpha} F_{\mu \alpha}\ .
\ear
From now on we will consider a radial electric field \( F_{01} = -F_{10} \) or a radial magnetic field \( F_{23} = -F_{32} \), such that the tensor \( F_{\mu\nu} \) becomes  compatible with spherical symmetry.  From the electrodynamics equations (\ref{eq-F}) we get, respectively,
\beq \label{FEFB}
 F^{10}=\frac{q_e}{r^2L_F}, \quad  F_{23}=q_m\sin\theta \to L_{F}^2 F=-\frac{q_{e}^2}{2r^4},F=\frac{q_{m}^2}{2r^4}
\eeq
where $q_e$ and $q_m$ represent, respectively, a constant electric and magnetic charge. The determination of the Lagrangian in the magnetic case is simpler because we can get $r(F)$, and if we can use Einstein's Equation to find $L(r)$, the problem is solved. The electric case is not so simple, since it involves $L_F$. To circumvent this, we will introduce an auxiliary object,$P_{\mu\nu}$,  such that \cite{Bronnikov:2000vy,Bronnikov:2017sgg, Bronnikov2023} 
\begin{equation}\label{auxiliary}
    P_{\mu\nu}=L_F F_{\mu\nu} \to  P = \frac{1}{4}P^{\mu\nu}P_{\mu\nu} = L_F^{2}F.
\end{equation}
With this
\begin{equation}\label{rP}
 P=-\frac{q_{e}^2}{2r^4}.
\end{equation}
and we can find $r(P)$. However, now, we also have to find $P(F)$ and, as we will see, it is not trivial. 

The stress-energy tensor (\ref{SET-F})  can be simplified to:
\bearr              \label{T-F}		
		{T^{EM}}^{\mu}_{\ \nu} 
		=  \diag\Big(L-\sqrt{FP},\ L-\sqrt{FP},\ L +\sqrt{FP},\ L +\sqrt{FP}\Big)\ . 
\ear   
  As pointed out above, we have that  
  ${T^{EM}}^{0}_{\ 0} = {T^{EM}}^{1}_{\ 1},\,{T^{EM}}^{2}_{\ 2} = {T^{EM}}^{3}_{\ 3} $, and has the same symmetries as the  Einstein tensor, Eq. (\ref{ET}). Below, we will consider the solution for each case separately. 

\subsection{Magnetic Charge as a Source}

  We will first consider the magnetic case ($q_e=0$) since it is the simplest. From Eq. \rf{FEFB}, we have $r(F)$. To find $L(r)$, we just need to use Eq. \rf{ET} to  obtain the expressions:
    \begin{equation}
        L = \frac{3\alpha^2 M^2}{r^6},\; F=\frac{q_{m}^2}{2r^4}\ .
    \end{equation}
Therefore, we get
\begin{equation}
  L=\frac{6\sqrt{2} \alpha^2 M^2}{q_m^3}F^{3/2}
\end{equation}

As expected, the above solution is well behaved as $\alpha\to0$, and we recover the Schwarzschild solution. In the next section, we will look for an electric source. 
%
  \subsection{Electric Charge as a Source}
Next, we consider $q_m=0$  to solve the above system. Similar to the previous case, we have $r(P)$ from Eq. (\ref{rP}) and, from the ${2\choose 2}$ component, we obtain  
    \begin{equation}
      L = -\frac{6\alpha^2 M^2}{r^6},\quad P=-\frac{q_{e}^2}{2r^4}
    \end{equation}
and
\begin{equation}
  L=\frac{6\sqrt{2} \alpha^2 M^2}{q_e^3} P^{\frac{3}{2}}.
\end{equation}
Now, if we find $P(F)$, the problem is solved. In order to achieve this,  note that the ${0\choose 0}$ and ${2\choose 2}$ components of the Einstein field equations (\ref{ET}) and \rf{eq-Ein} give us
\bearr              \label{T02}
  		G^{0}_{\ 0} - G^{2}_{\ 2}=\frac{f''(r)}{2}-\frac{f(r)}{r^2}+\frac{1}{r^2}= \kappa ^2T^{0}_{\ 0} - \kappa ^2T^{2}_{\ 2}
  		 = -\sqrt{FP}\ .
\ear 
and
\begin{equation}
      \sqrt{PF}=\frac{9 \alpha ^2 M^2}{r^{6}}.
\end{equation}
Since we have $r(P)$ we find
\begin{equation}
  L=-  \frac{2^{7/4}  \kappa ^3 q^{3/2}\alpha^2}{9  M}F^{3/4}
\end{equation}
As expected, the above solution is well behaved as $\alpha\to0$, and we recover the Schwarzschild solution. 

We thus conclude that the line element (\ref{eq1}) can be interpreted as generated by a nonlinear theory of electrodynamics with Lagrangian $L(F)\propto F^\eta$, where $\eta=3/2$ in the purely magnetic case and $\eta=3/4$ in the electric case.



\section{Static loop quantum black bounce} \label{sec_3}

The line element given in Eq. \rf{eq1} can be naturally extended to the region $r\to 0$, where a singularity exists. Given the electromagnetic nature of this space-time, it is easy to see that timelike geodesics never reach to the center, $r\to 0$, as they bounce off the potential barrier at some finite radius $r>0$. Radial null geodesics, instead, do reach the center, implying the existence of incomplete geodesics and, as a consequence, of a singularity.
 
By applying the Simpson-Visser prescription \( r \to \sqrt{r^2 + r_b^2} \) to the metric (\ref{eq1}), as described in \cite{Simpson:2018tsi}, a new spacetime that is well behaved on the whole real line, $r\in ]-\infty,+\infty[$, can be constructed. This procedure prevents the emergence of a singularity as \( r \to \sqrt{r^2 + r_b^2} \), similar to the approach in a LQG-inspired theory. As a result, we obtain
\begin{equation}\label{eq2}
    ds^2=\left[1\!-\!\frac{2M}{\sqrt{r^2+r_b^2}}\!+\!\frac{\alpha^2 M^2}{(r^2+r_b^2)^2}\right]\!dt^2 - \left[1\!-\!\frac{2M}{\sqrt{r^2+r_b^2}}+\frac{\alpha^2 M^2}{(r^2+r_b^2)^2}\right]^{-1}\!\!dr^2-(r^2+r_b^2) d\Omega^2.
\end{equation}
In principle, we treat $r_b$ as a free parameter. However, in certain cases, we define the bounce radius as a function of $\alpha$ according to Eq. (\ref{bounce-alpha}), without any loss of generality.
\begin{figure}[h!]
    \centering
    \includegraphics[scale = 0.45]{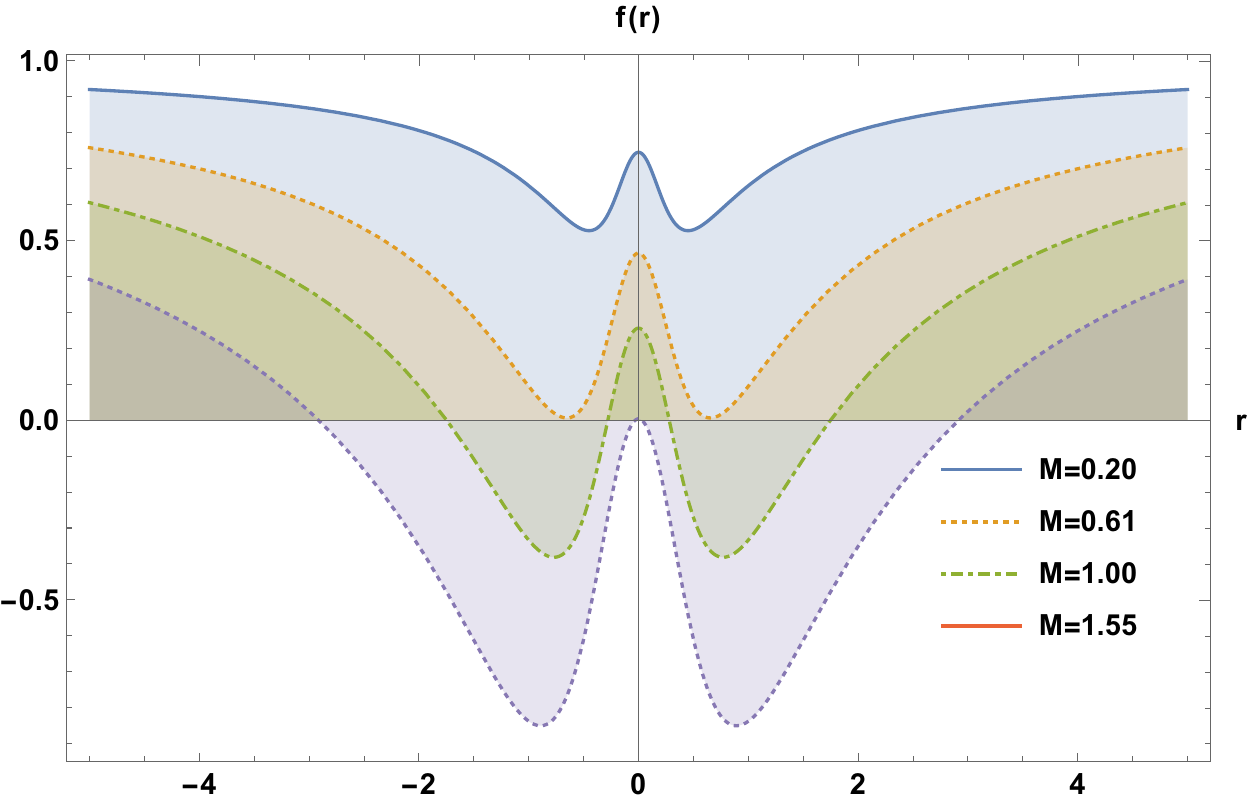}
    \caption{Plot of $g_{tt}$ related to the metric Eq. \rf{eq2}, for some values of the mass $M$, considering $\alpha=0.8$.}
    \label{Fig1}
\end{figure}

In Fig. \ref{Fig1}, we depict the metric coefficient $g_{tt}$ as a function of $r$ to verify the horizons' structure associated with the black bounce. It can be seen that the black bounce is a black hole with zero, two, or four horizons (zero, one, or two in the \( r > 0 \) domain). The black hole will be an extremal one, presenting two horizons (one in \( r > 0 \)),  if $M = \frac{4\alpha}{3\sqrt{3}}$. For $M$ below this value, the black bounce is horizonless and becomes a traversable wormhole.
\begin{figure}[h!]
    \centering
    \includegraphics[scale = 0.55]{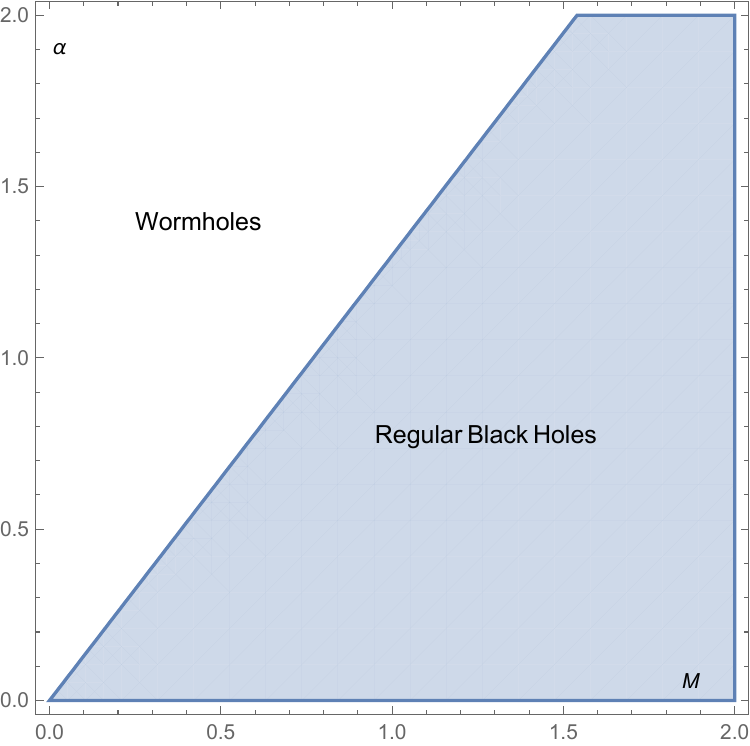}
    \caption{Parameter space $(M, \alpha)$ of the LQBB, highlighting the regions corresponding to wormholes (in white) and regular black holes (in blue).
}
    \label{Fig1.2}
\end{figure}
Fig. \ref{Fig1.2} illustrates the parameter space $(M, \alpha)$, highlighting the regions where wormholes and black holes can exist. The blue area indicates the presence of event horizons ($r_h\geq r_b$), corresponding to regular black holes, while the white area represents the region where LQBB is without horizons, indicating the occurrence of wormholes.

\subsection{Curvature}
\begin{figure}[h!]
    \centering
    \includegraphics[scale = 0.315]{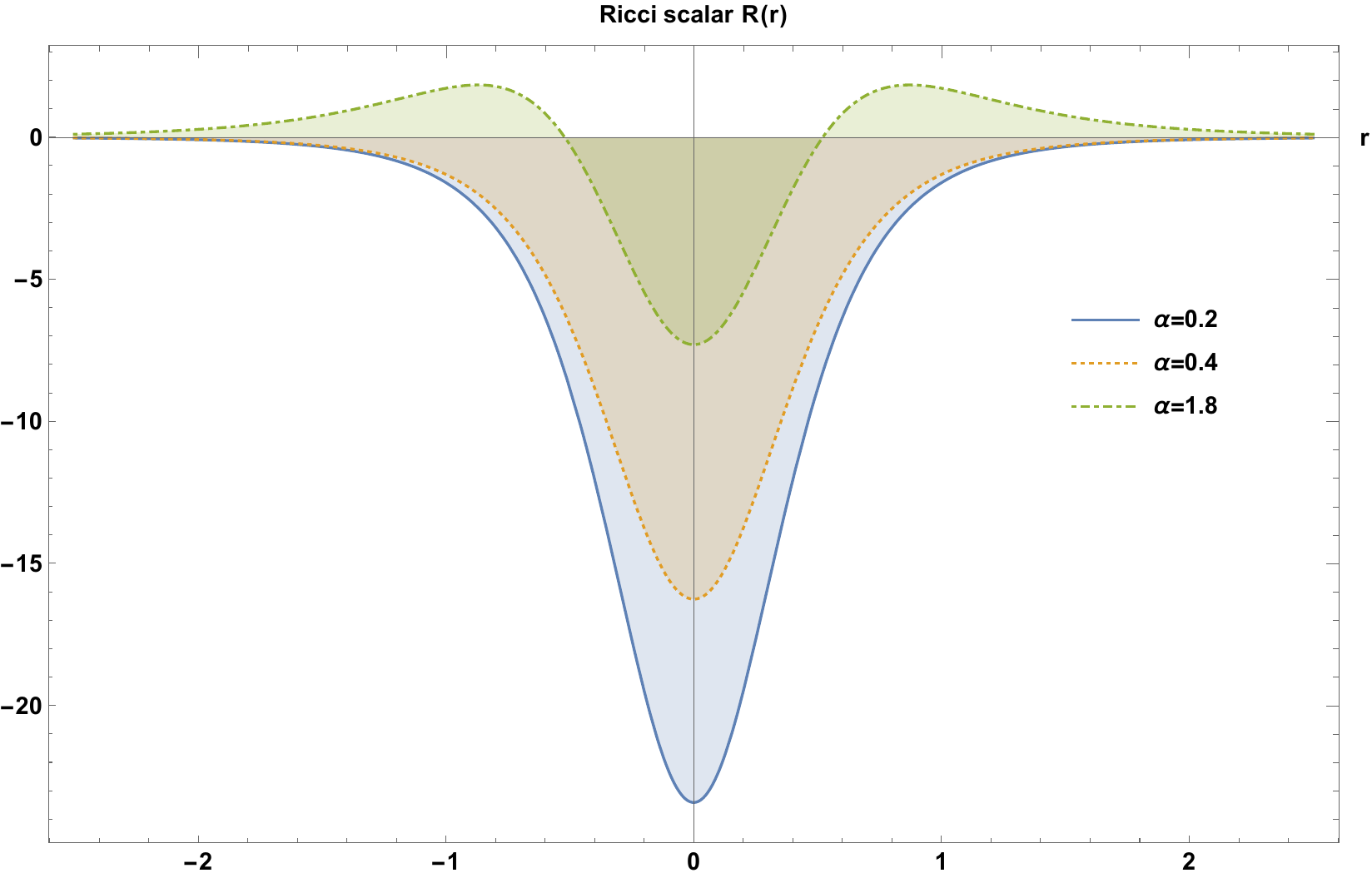  }
    \includegraphics[scale = 0.345]{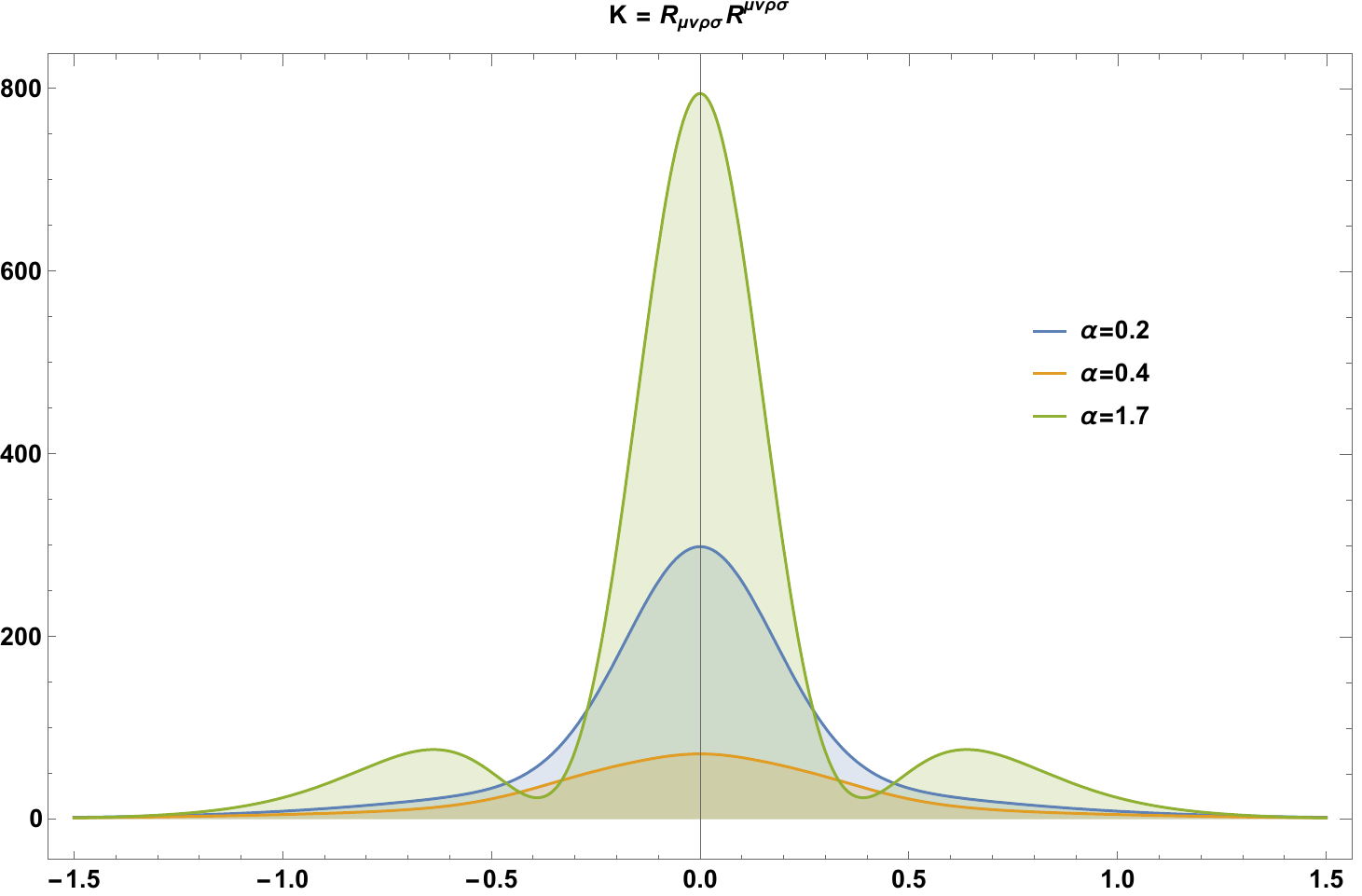}
    \caption{Plot of the Ricci and Kretschman scalars as a function of $r$, for some values of $\alpha$, considering $M=2$.}
    \label{Fig2}
\end{figure}
Using the line element (\ref{eq2}), we find that the Ricci scalar is given by
\begin{equation}\label{eq3}
 R=\frac{3\alpha^2 M^2 r^2 +r_b^2(r^2+r_b^2)^{\tmeio} \left(\sqrt{r_b^2+r^2}-3 M\right)}{\meio\left(r_b^2+r^2\right)^{4}} \ ,
\end{equation}
which indicates that the black bounce is regular in the entire domain of $r$. Fig. \eqref{Fig2} shows the regularity of LQBB, with both the Ricci and Kretschmann curvature scalars becoming more pronounced as the quantum parameter $\alpha$ increases near the bounce point ($r=0$).

\subsection{Energy Conditions}

Next, we examine the energy conditions and compare them with the results of section \ref{EC}. Using Einstein's equations to determine the stress-energy tensor that generates the metric  (\ref{eq2}), we find that the effective sources are characterized by
\begin{figure}[h!]
    \centering
    \includegraphics[scale = 0.60]{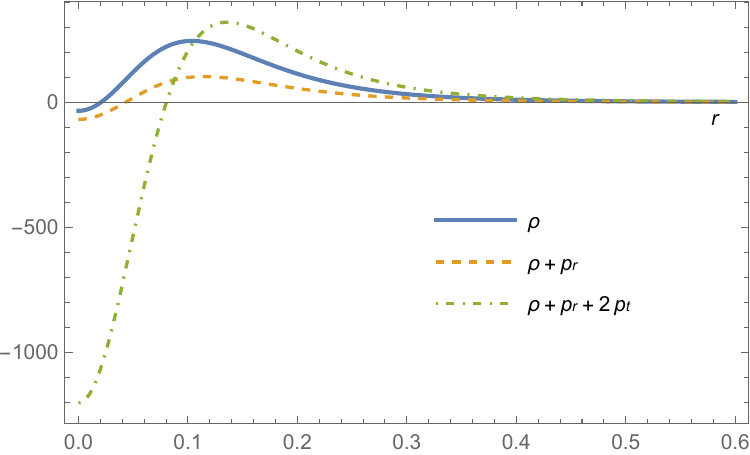}
\includegraphics[scale = 0.58]{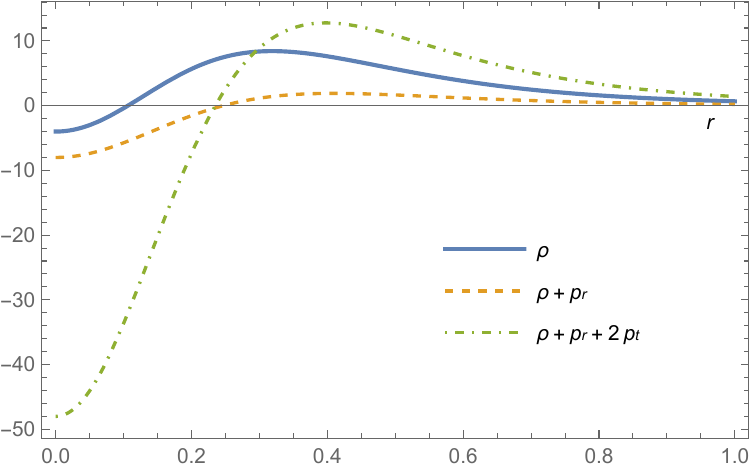}
    \caption{Energy density and its combinations with the pressures, considering $\alpha=0.1$ (left panel) and $\alpha=0.5$ (right panel), for $M=1$.
}
    \label{fig_cond1}
\end{figure}
\begin{eqnarray}
    \rho&=&\frac{3 \alpha ^2 M^2 r^2-r_bb^6-2 r_b^4 r^2-2 \alpha ^2 r_b^2 M^2-r_b^2 r^4}{\left(r_b^2+r^2\right)^4}+\frac{4 M r_b^2}{\left(r_b^2+r^2\right)^{5/2}},\nonumber\\
   p_r&=&-\frac{b^6+2 b^4 r^2+b^2 r^4+3 \alpha ^2 M^2 r^2}{\left(b^2+r^2\right)^4},\nonumber\\
   p_t&=&-\frac{-b^6-2 b^4 r^2+\alpha ^2 b^2 M^2-b^2 r^4-6 \alpha ^2 M^2 r^2}{\left(b^2+r^2\right)^4}-\frac{b^2 M}{\left(b^2+r^2\right)^{5/2}}.
\end{eqnarray}
In Fig. \ref{fig_cond1}, we present the energy density and its combinations with the pressures as functions of the radial coordinate, \(r\). The left panel corresponds to \(\alpha = 0.1\), while the right panel shows the results for \(\alpha = 0.5\) (\(M = 1\)). Notably, all energy conditions are violated in the region near the bounce point. Far away from it, all the conditions are met. However, as quantum gravity effects increase, represented by the parameter \(\alpha\), the severity of the violations diminishes. Thus, unlike the findings in the previous section, it becomes evident that some exotic sources are required when the regularization of the geometry is induced by the Simpson-Visser procedure. This conclusion will be further explored in the next section.

\section{Coupling with nonlinear electrodynamics and scalar field}
Before considering the specific case (\ref{eq2}), let us consider the more general one
\beq              \label{ds-cy}
    		ds^2=f(r)dt^2 - f(r)^{-1}dr^2 - \Sigma^2(r)\left(d\theta^2 + \sin^2\theta d\varphi^2\right).
\eeq
The corresponding components of the Einstein tensor are given by:
\bearr                 \label{GmnSV}
		G^0_0 = -\frac{f'(r) \Sigma '(r)}{\Sigma (r)}-\frac{f(r) \Sigma '(r)^2}{\Sigma (r)^2}-\frac{2 f(r) \Sigma ''(r)}{\Sigma
   (r)}+\frac{1}{\Sigma (r)^2},
\nnn 
		G^1_1 =-\frac{f'(r) \Sigma '(r)}{\Sigma (r)}-\frac{f(r) \Sigma '(r)^2}{\Sigma (r)^2}+\frac{1}{\Sigma (r)^2},				
\nnn
		G^2_2 = G^3_3 =-\frac{f'(r) \Sigma '(r)}{\Sigma (r)}-\frac{f''(r)}{2}-\frac{f(r) \Sigma ''(r)}{\Sigma (r)}.
\ear    
The first thing we must note is that 
\begin{equation}
 G^0_0-G^1_1=-\frac{2 f(r) \Sigma ''(r)}{\Sigma(r)}\neq 0
\end{equation}
Therefore, a NED source is not enough to generate this type of geometry. To address this, we will add a scalar field to the action (\ref{Action}), given by
\beq
    S_{\phi}=\int \sqrt{-g}d^4x\left[\epsilon g^{\mu\nu}\d_\mu \phi\d_\nu \phi
    		-V(\phi) \right]\ ,    			\label{Actionphi}
\eeq
where $\phi$ is the scalar field, $V(\phi)$ the scalar field potential, $\epsilon=\pm1$ depending on whether the scalar field is canonical ($+$) or phantom ($-$). Accordingly, we get the system of equations
\bearr
		\nabla_\mu \left[L_F F^{\mu\nu}\right]
			=0,\;\nabla_\mu^{*}  F^{\mu\nu}=0 ,
														\label{eq-FSV}
\\   
  \lal
      2\epsilon \nabla_\mu \nabla^\mu\phi =
		-  \frac{dV(\phi)}{d\phi}\ ,                      \label{eq-phi}
  \\\lal
     G_{\mu\nu}=R_{\mu\nu} - \frac 12 g_{\mu\nu} R = \kappa^2 \left(T_{\mu\nu} ^{\phi} + T_{\mu\nu} ^{EM}\right)\ ,			\label{eq-EinSV}
\ear
where
\begin{equation}
      T_{\mu\nu}^{\phi} =  2 \epsilon \d_\nu\phi\d_\mu\phi 
				    - g_{\mu\nu} \big (\epsilon \d^\alpha \phi \d_\alpha \phi - V(\phi)\big)\
\end{equation}
and $T_{\mu\nu} ^{EM}$ was given previously in Eq. (\ref{SET-F}). With the same assumptions as before, (\ref{eq-FSV}) implies 
\beq \label{FEFBSV1}
 F^{10}=\frac{q_e}{\Sigma^2L_F}, \quad  F_{23}=q_m\sin\theta \to L_{F}^2 F=-\frac{q_{e}^2}{2\Sigma^4},F=\frac{q_{m}^2}{2\Sigma^4}.
\eeq
By using the auxiliary field (\ref{auxiliary}) we get
\begin{equation}
     P=-\frac{q_{e}^2}{2\Sigma^4},F=\frac{q_{m}^2}{2\Sigma^4}.
\end{equation}
Now, assuming that $\phi = \phi(r)$, we get for the stress-energy tensors :
\bearr         \label{T-phiSV}
		{T^{\phi}}^{\mu}_{\ \nu} = \epsilon f\phi'{}^2 \diag (1, -1, 1, 1)
		 + \delta^\mu _{\nu} V(\phi)\ , 
\yyy           \label{T-FSV}		
		{T^{EM}}^{\mu}_{\ \nu} 
		=  \diag\Big(L-\sqrt{PF},\ L-\sqrt{PF},\ L +\sqrt{PF},\ L +\sqrt{PF}\Big)\ . 
\ear   
Note that now we have ${T}^{0}_{\ 0} \neq {T}^{1}_{\ 1},\,{T}^{2}_{\ 2} = {T}^{3}_{\ 3} $, with the same symmetries as the  Einstein tensor, Eq. (\ref{GmnSV}). Before considering each case separately, we will use $\Sigma^2=r^2+r_b^2$ and simplify further by finding the solution to the scalar field.  

 The scalar field equation \eqref{eq-phi} yields
\beq
	\frac{2}{\Sigma^2(r)}\partial_{r}\left[\Sigma^2(r)f(r)\partial_{r}\phi\right]=V'(\phi)\ .
			\label{eq-phi-2}
\eeq
  Furthermore, by combining again the ${0\choose 0}$ and ${1\choose 1}$ components of 
  the Einstein field equations \rf{eq-Ein} as
\beq        \label{T01}
		G^{1}_{\ 1} - G^{0}_{\ 0} = \frac{2 f \Sigma ''}{\Sigma} 
		= \kappa ^2T^{1}_{\ 1} - \kappa ^2T^{0}_{\ 0}=- 2 \kappa ^2 \epsilon  f \phi '^2 \ ,
\eeq   
we find that the scalar field is given by
\begin{equation}
 \phi '^2= \frac{ \Sigma ''}{\kappa^2\Sigma}=\frac{ r_b^2}{\kappa^2\Sigma^4}\to  \phi(r)=\frac{1}{\kappa}\arctan{\left(\frac{r}{r_b}\right)},
\end{equation}
and the potential $V(r)$ is 
\begin{equation}
V(r)=-\frac{4 M r_b^2}{\kappa ^2} \left[\frac{\alpha ^2 M}{4 \left(r_b^2+r^2\right)^4}-\frac{1}{5 \left(r_b^2+r^2\right)^{5/2}}\right]=-\frac{4 M r_b^2}{\kappa ^2} \left[\frac{\alpha ^2 M}{4 \Sigma^8}-\frac{1}{5 \Sigma^{5}}\right].
\end{equation}
\begin{figure}[h!]
    \centering
    \includegraphics[scale = 0.30]{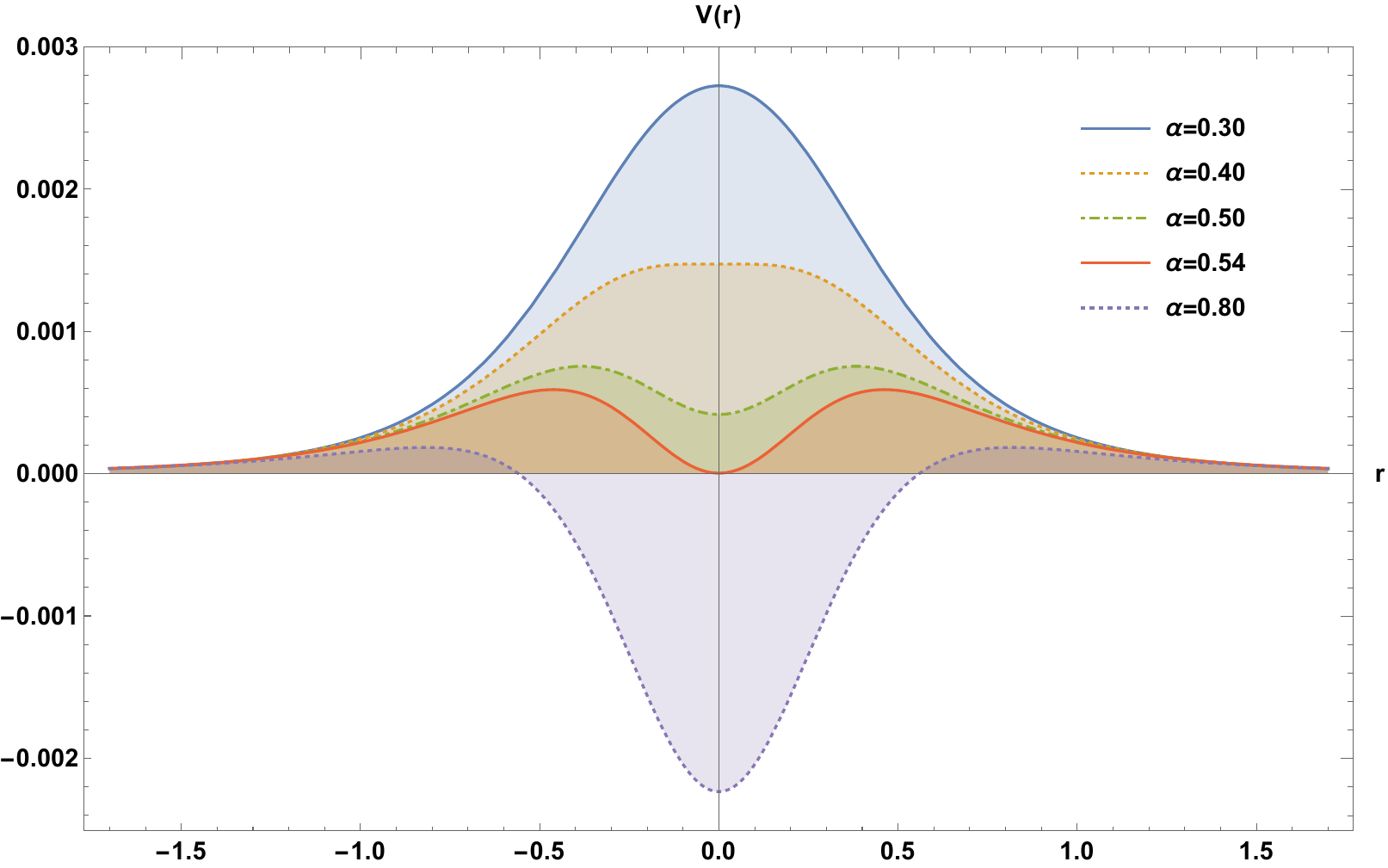}
 \includegraphics[scale = 0.642]{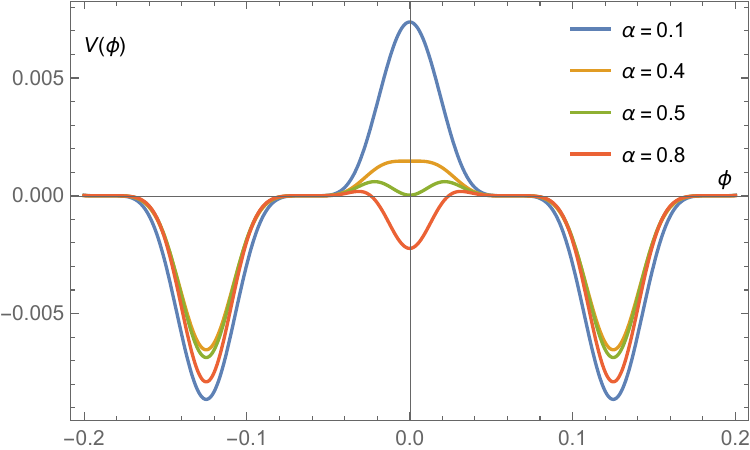}
    \caption{Left: Potential associated with the scalar field as a function of $r$. Right: The same quantity as a function of $\phi$.}
    \label{fig_pot}
\end{figure}
In Fig. \eqref{fig_pot}, the left side shows that the height of the potential barrier decreases as quantum effects increase (i.e., as \(\alpha\) increases). This reduction in barrier height facilitates the tunneling of a quantum particle to another region of the universe. A similar trend is observed on the right side of the figure. Additionally, the potential at the origin results in a static equilibrium state for the scalar field when \(\alpha > 2^{-4/3} \approx 0.4\). It is worth mentioning that in the limit $\alpha\to 0$ one recovers the results discussed in \cite{Rodrigues:2023vtm}. 

Now, the stress tensor of the scalar field simplifies to 
\begin{equation}
    {T^{\phi}}^{\mu}_{\ \nu} = - (1-\frac{2M}{\Sigma}+\frac{\alpha^2 M^2}{\Sigma^4})\frac{ r_b^2}{\kappa^2\Sigma^4} \diag (1, -1, 1, 1)
		 - \delta^\mu _{\nu} \-\frac{4 M r_b^2}{\kappa ^2} \left(\frac{\alpha ^2 M}{4 \Sigma^8}-\frac{1}{5 \Sigma^{5}}\right)
\end{equation}
With this, we will now consider the magnetic and electric cases separately. 

\subsection{Magnetic Charge as a Source}
We will now find magnetically charged NED and scalar field sources for LQBB. Since we already have $\Sigma(F)$, we need to find $L(\Sigma)$. For this, we just need the ${0\choose 0}$ component of the Einstein Equation. First, note that the ${0\choose 0}$ component of the Einstein tensor can be written as
\begin{equation}
  G^{0}_{\ 0}=-5\frac{\alpha ^2 M^2 r_b^2}{\Sigma^8}+3\frac{\alpha ^2 M^2 }{\Sigma^6}-\frac{4 M r_b^2}{\Sigma^{5}}+\frac{r_b^2}{\Sigma^4}.
\end{equation}
On the other hand, the right-hand side is given by
\begin{equation}
  {(T^{\phi})}^{0}_{\ 0} +{(T^{EM})}^{0}_{\ 0} =L- (1-\frac{2M}{\Sigma}+\frac{\alpha^2 M^2}{\Sigma^4})\frac{ r_b^2}{\kappa^2\Sigma^4}- \-\frac{4 M r_b^2}{\kappa ^2} \left(\frac{\alpha ^2 M}{4 \Sigma^8}-\frac{1}{5 \Sigma^{5}}\right)
\end{equation}
Therefore, we get directly
\begin{equation}
L(F)=\frac{6\sqrt{2}M^2\alpha^2}{q_m^3\kappa ^2}F^{3/2}+r_b^2\frac{3 M 2^{9/4} q^{5/2}}{5 q_m^{1/2} \kappa ^2}F^{5/4}.
\end{equation}
As expected, when we take the limit $r_b\to0$ we recover the LQG black hole. We also can see that, in the limit $\alpha\to0$ we recover the source for the regularized Scharzschild case \cite{Bronnikov:2021uta}. Note that the effective NED source is directly influenced by the quantum properties of spacetime determined by LQG, which are encoded in the parameter $\alpha$. 

\subsection{Electric Charge as a Source}
Now we need the ${2\choose 2}$ component of the Einstein tensor, given by
\begin{equation}
  G^{2}_{\ 2}=  -\frac{7\alpha ^2 M^2r_b^2}{\Sigma^8} +\frac{6\alpha ^2 M^2}{\Sigma^6}-\frac{M r_b^2 }{\Sigma^5}+\frac{ r_b^2 }{\Sigma^4}.
\end{equation}
Since $T^{0}_{\ 0}=T^{2}_{\ 2}$ we get
\[
\kappa^2L = -\frac{5\alpha ^2 M^2r_b^2}{\Sigma^8} + \frac{6\alpha ^2 M^2}{\Sigma^6} - \frac{19 M r_b^2 }{5\Sigma^5} + \frac{2r_b^2 }{\Sigma^4},
\]
and
\[
\kappa^2L = 
- \frac{20\alpha^2 M^2 r_b^2 P^2}{q_e^4} 
+ \frac{6\alpha^2 M^2 (2P)^{3/2}}{q_e^{3}} 
- \frac{19 M r_b^2 (2P)^{5/4}}{q_e^{5/2}} 
+ \frac{4 r_b^2 P}{q_e^2}.
\]

Now, we need to find $P(F)$. We have
\begin{equation}
    G^{2}_{\ 2} - G^{0}_{\ 0} = \kappa ^2T^{2}_{\ 2} - \kappa ^2T^{0}_{\ 0}
  		 = \sqrt{PF}\ 
\end{equation}
and therefore we get
\begin{equation} 
\frac{8 \alpha^2 M^2 r_b^2 P^2}{q_e^4} - \frac{6 \sqrt{2} \alpha^2 M^2 P^{3/2}}{q_e^3} - \frac{3 \cdot 2^{5/4} M r_b^2 P^{5/4}}{q_e^{5/2}}=-\frac{\kappa ^2 q^2}{L_F \Sigma ^4}=\sqrt{PF}
\end{equation}
or
\begin{equation} 
\frac{8 \alpha^2 M^2 r_b^2 P^{3/2}}{q_e^4} - \frac{6 \sqrt{2} \alpha^2 M^2 P}{q_e^3} - \frac{3 \cdot 2^{5/4} M r_b^2 P^{1/4}}{q_e^{5/2}}=\sqrt{F}
\end{equation}
Therefore, for $r\neq 0$, it is not possible to find $P(F)$. However, for $r_b=0$ we recover the LQG black hole.
\section{Thermodynamics}

Let's calculate the surface gravity at the event horizon for the regular black hole case. The Killing vector, which is null at the event horizon, is $\xi^\mu = \partial_t$. This gives the following norm:
\begin{equation}
    \xi^\mu \xi_\mu = g_{\mu\nu} \xi^\mu \xi^\nu = g_{tt} = - \left[1 - \frac{2M}{\sqrt{r^2 + r_b^2}} +\frac{\alpha^2 M^2}{(r^2+r_b^2)^2}\right]. 
\end{equation}

Then, we have the following relation for the surface gravity $\kappa$ (see for instance \cite{Cembranos2017, Balakin2016}):
\begin{equation}
    -\nabla_\nu g_{tt} = 2\kappa \xi_\nu,
\end{equation}
from which the Hawking temperature is calculated as follows:
\begin{equation}
T_H=\frac{\kappa}{2\pi}=\frac{1}{4\pi}\left.\frac{d g_{tt}}{dr}\right|_{r=r_h}.
\end{equation}
\begin{figure}[h!]
\centering
\includegraphics[width=0.47\textwidth]{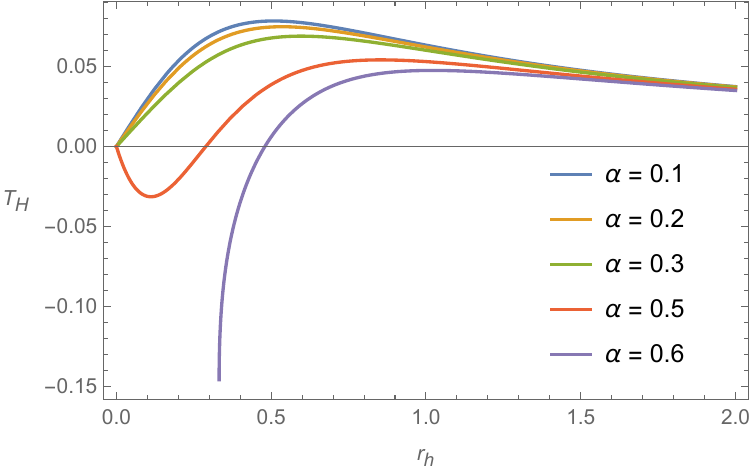}
\includegraphics[width=0.495\textwidth]{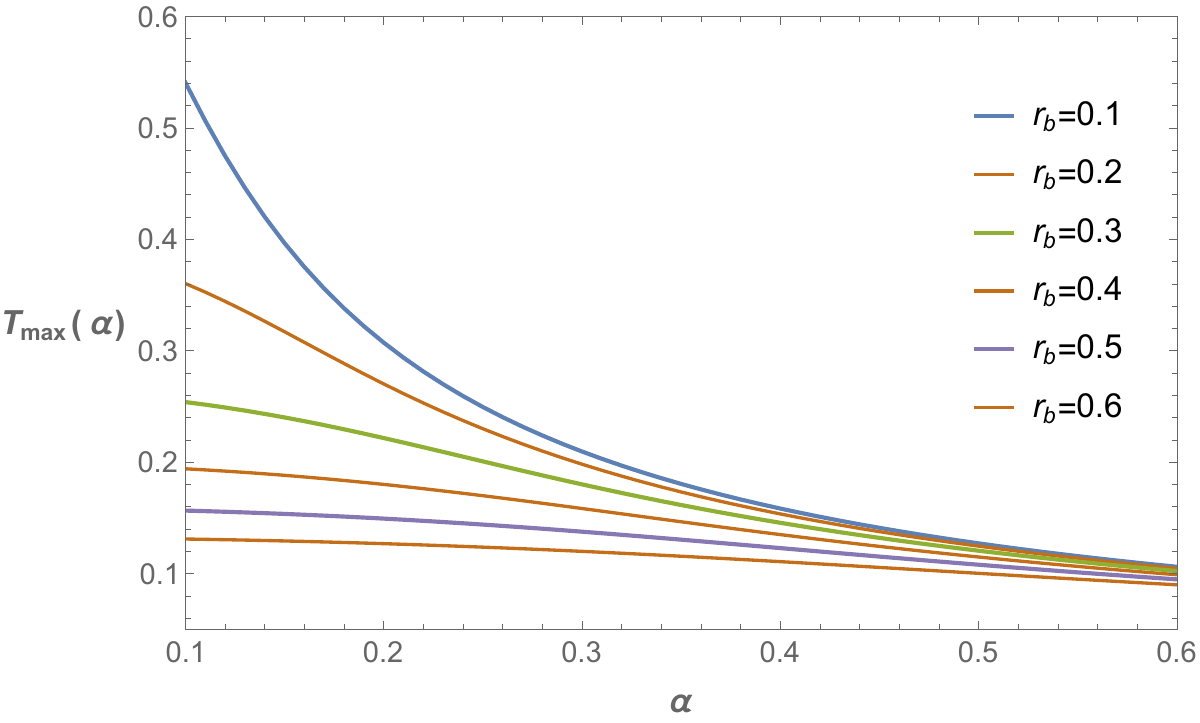}
\caption{Left panel: Hawking temperature as a function of $r_h$, for different values of the quantum parameter $\alpha$, considering $r_b=0.5$. Right panel: Local maxima of the Hawking temperature as a function of $\alpha$, varying the bounce parameter, $r_b$.}
\label{fig:HawkingTemperature}
\end{figure}

As the expression to find the horizon radius from $g_{tt}=0$ is very involved, we write the value of the black hole mass as a function of the horizon radius, given by
\begin{equation}
M=\frac{(r_h^2+r_b^2)^2-\sqrt{\left(r_b^2+r_h^2\right)^3 \left(r_b^2+r_h^2-\alpha ^2\right)}}{\alpha ^2 \sqrt{r_b^2+r_h^2}}.\label{MassBH}
\end{equation}
Hence, it follows that the temperature of Hawking radiation for our regular black hole is given by
\begin{equation}
    T_H=\frac {r_h}{2 \pi \alpha^2} \left[1 - \sqrt{1-\frac{\alpha^2}{r_b^2 + r_h^2}} \right] \left[2\sqrt{1 - \frac{\alpha^2}{r_b^2 + r_h^2}}-1\right].\label{HawkTemp1}
\end{equation}
The above expression shows that the temperature is null for 
\begin{equation}\label{T0}
r_h=\frac{1}{\sqrt{3}}\sqrt{4 \alpha ^2-3 r_b^2}
\end{equation}\label{ctemp}
For a radius fewer than this, the temperature becomes negative. Therefore, we have a remnant  unless $\alpha>\sqrt{3}r_b/2$, namely if the LQG parameter is higher than the regularization one.  In other words, the spacetime quantum effects favour the formation of remnants. Below we will analyze if this is a stable point. Fig. \ref{fig:HawkingTemperature} clearly reveals this behavior. In the left panel, we have the Hawking temperature as a function of the event horizon radius, $r_h$, for a loop quantum black bounce, according to Eq. (\ref{HawkTemp1}). The quantum parameter $\alpha$ is varied across different values, and for $r_b=0.5$, the critical value is $\alpha_c\approx 0.43$. In the right panel, we depict the variation of the maxima of the Hawking temperature with $\alpha$, for different bounce parameters, $r_b$. Observe that these maxima decrease with increasing of both the parameters.

Note that the expression for the Hawking temperature can also be written as
\begin{equation}
    T_H=\frac{1}{4\pi r_h}\left[1-\frac{r_b^2}{r_h^2}-\frac{3\alpha^2}{4r_h^2}+\frac{3 r_b^2\alpha^2}{2r_h^4}+ \mathcal{O}(\alpha^4,r_b^4)\right],
\end{equation}
where the multiplicative factor is just the Hawking temperature for the Schwarzschild black hole, which is obtained when $(\alpha, r_b) \to 0$. 
                    
Another thermodynamic quantity of interest is the entropy, $S$, which we calculate via
\begin{equation}
    dS=\frac{dM}{T_H}.\label{entropy1}
\end{equation}
Derivating Eq. (\ref{MassBH}) relative to $r_h$, dividing by the expression for the Hawking temperature (\ref{HawkTemp1}) and integrating again, we arrive at
\begin{equation}
    S=\pi  \left[r_h \sqrt{r_b^2+r_h^2-\alpha ^2}+\left(\alpha ^2+r_b^2\right) \log \left(\sqrt{r_b^2+r_h^2-\alpha ^2}+r_h\right)\right].
\end{equation}
The first thing we can see is that at the  radius of null temperature, given by Eq. (\ref{T0}), we get
\begin{equation}
    S=\pi  \frac{\alpha}{3}\left[\sqrt{4 \alpha ^2-3 r_b^2} +4\alpha \log \left(\frac{2\alpha+\sqrt{4 \alpha ^2-3 r_b^2}}{\sqrt{3}}\right)\right].
\end{equation}
Therefore, as expected, we have a finite entropy as a remnant. Expanding this in $\alpha$ and $r_b$ we have
\begin{equation}
    S=\pi r_h^2 \left[1 + \frac{r_b^2}{2 r_h^2} + \frac{\alpha^2 r_b^2}{4 r_h^4} - \frac{\alpha^2}{2 r_h^2}\right]+\pi\left(r_b^2 + \alpha^2\right) \log(2 r_h)+\mathcal{O}(\alpha^4,r_b^4),
\end{equation}
where the first multiplicative factor represents the expression for the entropy of a Schwarzschild black hole. Note that the logarithmic corrections to the entropy arise solely due to the quantum nature of spacetime, within the framework of LQG that we are considering.
\begin{figure}[h]
    \centering
    \includegraphics[scale = 0.64]{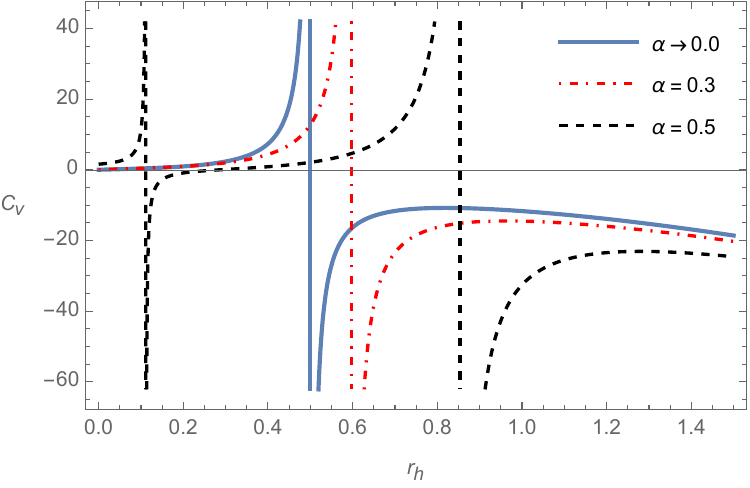}
    \includegraphics[scale = 0.4]{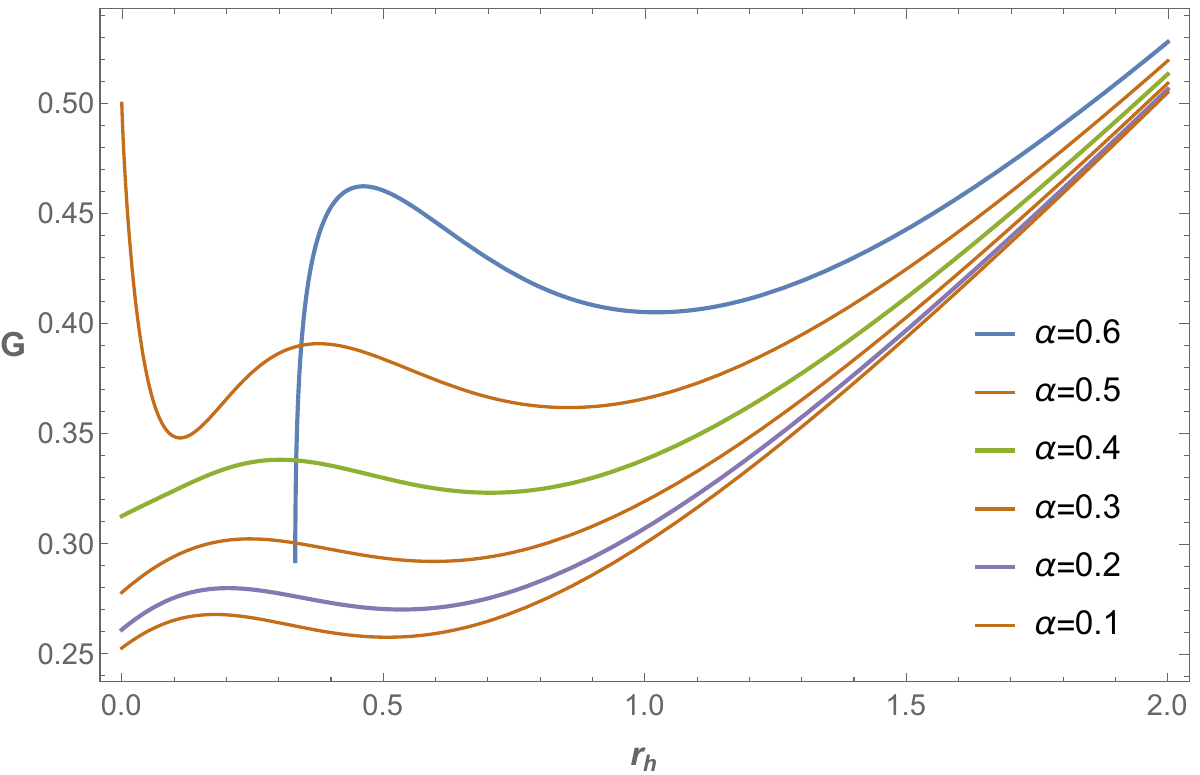}
     \caption{Left panel: Thermal capacity as a function of $r_h$, for $r_b=0.5$, considering some values of $\alpha$. Right panel: Gibbs free energy as a function of $r_h$, for several values of $\alpha$ and $r_b=0.5$.}
    \label{capgibbs}
\end{figure}
In order to analyze the phase structure of the black holes under consideration, we will calculate the thermal capacity at constant volume, $C_v$, and the Gibbs free energy, $G$, given respectively by
\begin{equation}
    C_v=\frac{dM}{dT_H}=\frac{dM/dr_h}{dT_H/dr_h};\text{ } G= M-TS.
\end{equation}
\begin{figure}[h!]
     \centering
     \includegraphics[scale = 0.5]{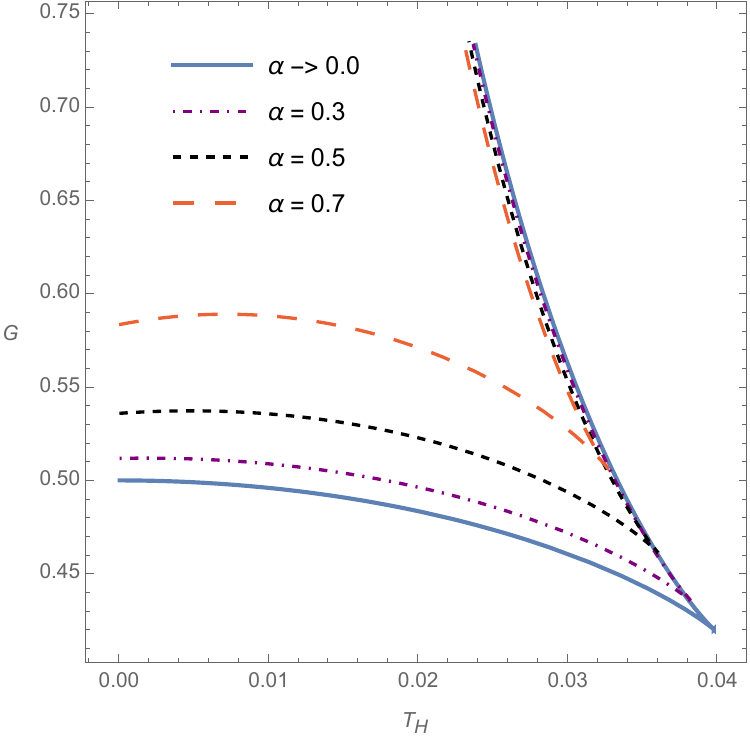}
     \includegraphics[scale = 0.5]{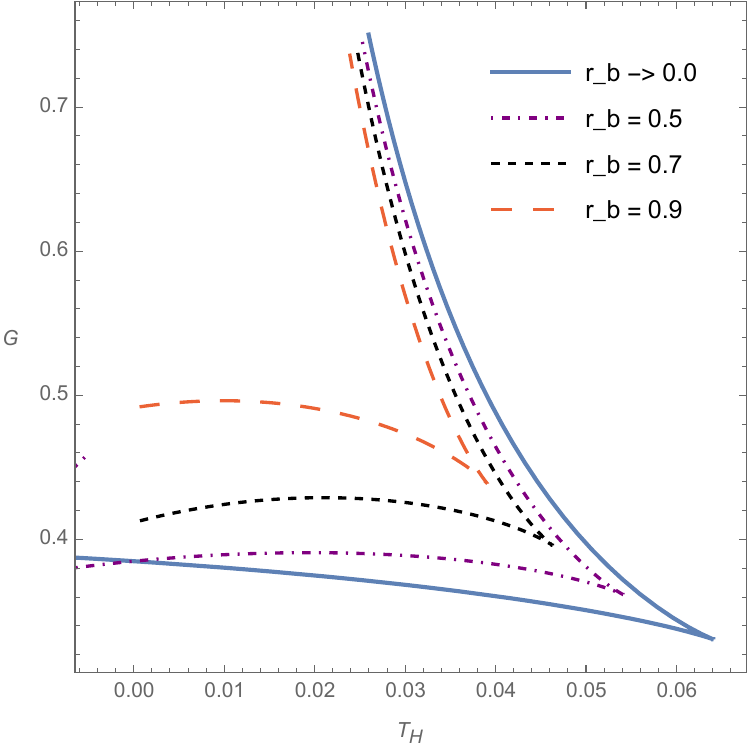}
     \caption{Gibbs free energy as a function of Hawking temperature $T_h$ for several values of $\alpha$ ($r_b=0.5$, left) and several values of $r_b$ ($\alpha=0.5$, right).}
    \label{gibbs_Th} 
\end{figure}
%
In Fig. \ref{capgibbs}, we depict \(C_v\) (left panel) and \(G\) (right panel) as functions of the horizon radius \(r_h\), which serves as an order parameter of the system, for $r_b=0.5$. The behavior of \(C_v\) reveals local phase transitions of first order, which are characterized by discontinuities where \(C_v\) shifts from positive (indicating local stability) to negative (local instability) values. Notably, for \(\alpha = 0.5\), we observe two first-order phase transitions and one second-order phase transition, marked by a smooth change in the sign of \(C_v\). This complex behavior is due to the influence of the LQG parameter \(\alpha\), which alters the thermodynamic landscape, leading to such multiple phase transitions. In fact, there exists a specific range of \(\alpha\) values where these phenomena are observed, underscoring the critical impact of ($\alpha$) on the phase structure of the system.

Regarding the Gibbs free energy, the plot in the right panel of Fig. \ref{capgibbs} reveals the absence of phase transitions. However, the system is globally thermodynamically unstable due to the positive values of \(G\) for any \(r_h\). Notably, at a critical value around $\alpha=0.5$, the plot reveals two local minima in the Gibbs free energy. This indicates the presence of two stable black hole phases with different sizes. On the other hand, on evaluating the dependence of $G$ on $T_H$, Fig. \ref{gibbs_Th} illustrates the coexistence of two distinct thermodynamic phases. At low temperatures, different values of \(\alpha\) and $r_b$ lead to distinct phases, with the differences becoming less pronounced as \(\alpha,r_b\) increases. At high temperatures, the curves converge, indicating a universal phase at lower critical temperatures for increasing ($\alpha$, $r_b$).
%
\section{Stationary loop quantum black bounce}
Applying the revised Newman-Janis algorithm \cite{Azreg-Ainou:2014pra,Azreg-Ainou:2014nra} to the metric (\ref{eq2}), we obtain the Kerr-like LQG-inspired black bounce solution
\begin{equation}
   \!\!ds^2=\frac{\Psi}{\rho^2} \left\lbrace\frac{\Delta}{\rho^2}(dt-a\sin^2{\theta}d\phi)^2-\frac{\rho^2}{\Delta}dr^2-\rho^2d\theta^2-\frac{\sin^2{\theta}}{\rho^2}\left[a dt-\left(r^2+r_b^2+a^2\right)d\phi\right]^2 \right\rbrace, \label{rotmetr}
\end{equation}
where 
\begin{eqnarray}
\Delta&=&(r^2+r_b^2)\left[1-\frac{2M}{\sqrt{r^2+r_b^2}}+\frac{\alpha^2 M^2}{(r^2+r_b^2)^2}\right]+a^2,\text{ and}\label{eqDelta}\\
\rho^2&=& r^2+r_b^2+a^2\cos^2{\theta}.
\end{eqnarray}
The function  $\Psi=\Psi(r,\theta,a)$ must satisfy $G_{r\theta}=0$ in addition to Einstein's equations \cite{Azreg-Ainou:2014aqa}. Considering our solution, we find that $\Psi=\rho^2$. 

\begin{figure}[h!]
    \centering
    \includegraphics[scale = 0.70]{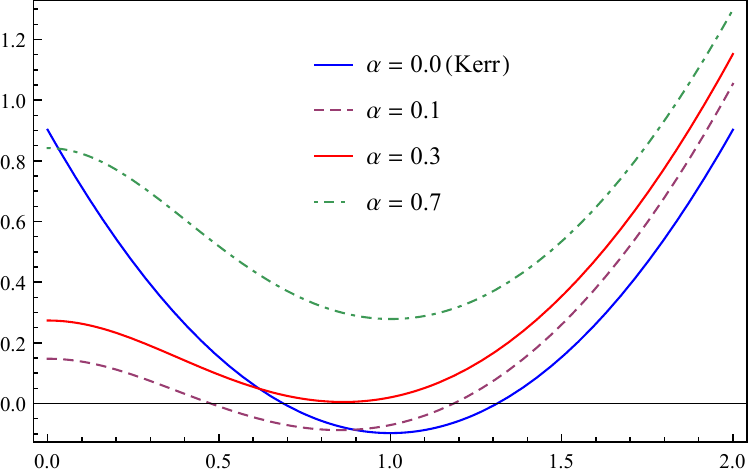}
    \caption{Plot of $\Delta(r)$, for some values of $\alpha$, considering $M=1$ and $a=0.95$.}
    \label{Fig3}
\end{figure}

The horizons are determined by the condition $\Delta=0$. Fig. \ref{Fig3} illustrates the scenarios where there are zero, one, or two horizons within the domain $r\geq0$. 

\begin{figure}[h!]
    \centering
    \includegraphics[scale = 0.66]{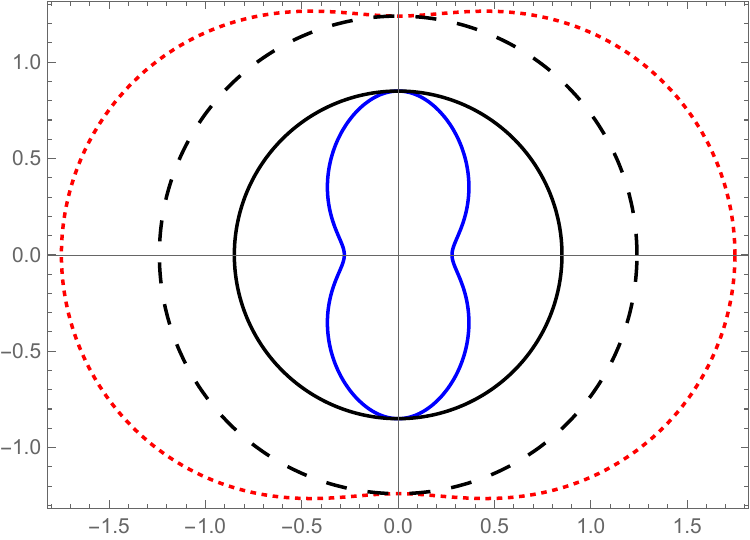}
  \includegraphics[scale = 0.56]{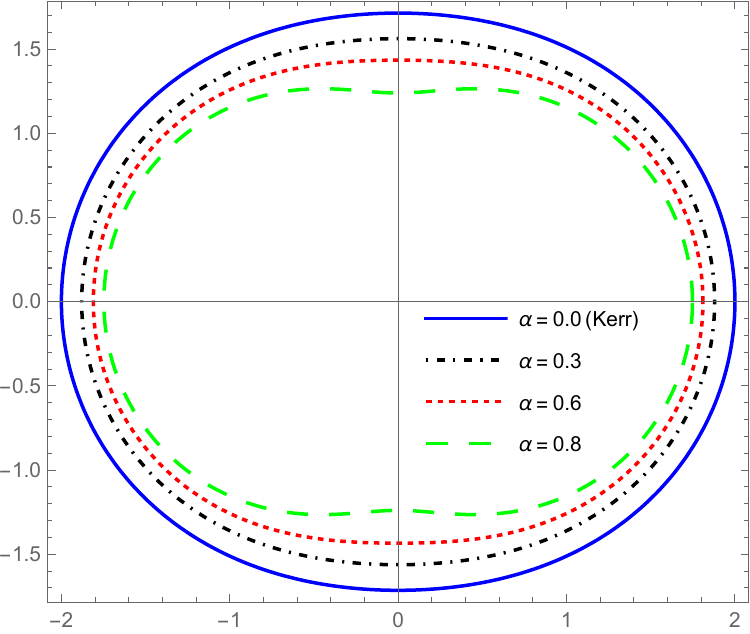}
    \caption{Left panel: Internal ergosphere (solid, blue), internal horizon (solid, black), external horizon (long-dashed, black), and external ergosphere (dashed, red), for $a=0.7$ and $\alpha=0.8$. Right panel: External ergospheres, varying $\alpha$, for $a=0.7$.}
    \label{ergo}
\end{figure}
In Fig. \eqref{ergo}, we depict the profiles of the ergospheres, obtained by solving \(g_{tt} = 0\). On the left side, we depict the structure of the black bounce, with the internal and external horizons and ergospheres; on the right side, we present the external ergospheres while varying \(\alpha\) and fixing \(a\), including the Kerr black hole (for which $\alpha=0$). Note that the size of the ergospheres decreases with an increase in the LQG parameter \(\alpha\). This reduction occurs because higher values of \(\alpha\) introduce significant quantum corrections, shrinking and deforming the region where particles are forced to co-rotate with the black hole.

Concerning the curvature of the rotating LQBB spacetime, it can be shown that the Ricci scalar \( R \) is generally given by $R = \frac{N}{D}$, where \( N \) depends on the metric functions appearing in (\ref{rotmetr}) and their derivatives up to second order, while \( D = 2 \rho^6 \Psi^3 \) \cite{Azreg-Ainou:2014aqa}. Therefore, the rotating LQBB spacetime would exhibit a singularity provided $r^2 + r_b^2 + a^2 \cos^2 \theta = 0$. However, this quantity cannot vanish because \( r_b \neq 0 \). To better understand the geometry of this region, let us consider the line element of Eq.(\ref{rotmetr}) in the region $r=0$, where it can be recast as 
\begin{eqnarray}\label{eq:ds2r0}
    ds^2=dt^2-\frac{\left(2Mr_b-\alpha^2 M^2/r_b\right)}{r_b^2+a^2\cos^2\theta}\left(dt-a\sin^2\theta d\phi\right)^2-(r_b^2+a^2\cos^2\theta)d\theta^2-(r_b^2+a^2)\sin^2\theta d\phi^2
\end{eqnarray}
For the sake of comparison, one can consider the usual Kerr scenario, the limit $\alpha\to 0$ and $r_b\to 0$, for which the above line element boils down to 
\begin{eqnarray}
    ds^2=dt^2-a^2\cos^2\theta d\theta^2-a^2\sin^2\theta d\phi^2=dt^2-d\tilde{r}^2-\tilde{r}^2d\phi^2 \ ,
\end{eqnarray}
where we have defined $\tilde{r}=a\cos\theta$, such that $|\tilde{r}|\in [0,a]$. Thus, this line element represents the geometry of a flat $2+1$ dimensional circle of radius $a$ in Minkowski space-time in polar coordinates. The problem comes when we consider the case $\theta=\pi/2$ (equatorial plane of the Kerr geometry), because there $\tilde{r}=0$ represents a circumference of vanishing radius (at constant time) on which the Kerr curvature scalars diverge. In our case, the last two terms of the line element (\ref{eq:ds2r0}) represent the geometry of a rotating ellipsoid of radius $r_b$ in the $z$ direction, and of radius $\sqrt{r_b^2+a^2}$ on the $x-y$ plane.  As one can see, this ellipsoid degenerates into two circles when $r_b\to 0$ (one of those circles corresponds to the interval $\tilde{r}\in [0,a]$, where $z>0$, while the other is in $\tilde{r}\in [-a,0]$, where $z<0$). This ellipsoid represents the transit (or bounce) region, with the parameter $a$ controlling its deformation from a sphere in the rotating case. As mentioned before, the curvature invariants are regular on this ellipsoid, including the equatorial circumference $\theta=\pi/2$. Further details on the properties of this ellipsoid and its corresponding Kerr-Schild coordinates will be given elsewhere.

\begin{figure}
    \centering
    \includegraphics[scale = 0.6]{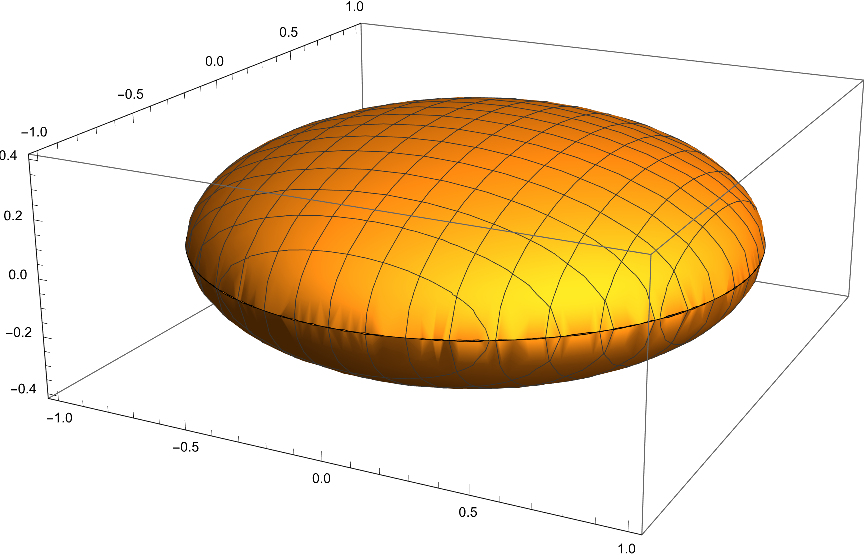}
    \caption{Ellipsoidal transit surface $r=0$ of the rotating LQBB  for $a=0.9$ and $r_b=0.4$.}
    \label{torus}
\end{figure}

\subsection{Shadow of Rotating Loop Quantum Black Bounces}

To determine the contour of a black hole (BH) shadow, we will follow \cite{Jusufi:2020odz}. Thus, we start with the Hamilton-Jacobi equation, given by
\begin{equation}
\frac{\partial S}{\partial \sigma} = -\frac{1}{2} g^{\mu\nu} \frac{\partial S}{\partial x^\mu} \frac{\partial S}{\partial x^\nu},
\end{equation}
where $\sigma$ is the affine parameter and $S$ is the Jacobi action. To find a separable solution, we express the action in terms of known constants of motion as:
\begin{equation}
S = \frac{1}{2} \mu^2 \sigma - Et + J \phi + S_r(r) + S_\theta (\theta),
\end{equation}
where $\mu$ is the mass of the test particle, $E = -p_t$ is the conserved energy and $J = p_\phi$ is the conserved angular momentum (concerning the symmetry axis). For a photon, $\mu = 0$. These equations yield the following equations of motion (see, for instance, \cite{Toshmatov2017}):
\begin{equation}
\Sigma \frac{dt}{d\sigma} = \frac{r^2 + a^2}{\Delta} \left[E(r^2 + a^2) - a J \right] - a \left(a E \sin^2 \theta - J\right),
\end{equation}
\begin{equation}
\Sigma \frac{d\phi}{d\sigma} = \frac{a}{\Delta} \left[E(r^2 + a^2) - a J\right] - \left(a E - \frac{J}{\sin^2 \theta}\right),
\end{equation}
\begin{equation}
\Sigma \frac{dr}{d\sigma} = \pm \sqrt{R(r)},
\end{equation}
\begin{equation}
\Sigma \frac{d\theta}{d\sigma} = \pm \sqrt{\Theta(\theta)},
\end{equation}
where
\begin{equation}
R(r) = \left[X(r)E - a J\right]^2 - \Delta(r) \left[K + (J - aE)^2 \right],
\end{equation}
\begin{equation}
\Theta(\theta) = K + a^2 E^2 \cos^2 \theta - J^2 \cot^2 \theta,
\end{equation}
with $X(r) = (r^2 + a^2)$. The function $\Delta(r)$ is defined by Eq. \eqref{eqDelta}, while $K$ is the Carter separation constant.
\begin{figure}[h!]
    \centering
    \includegraphics[scale = 0.35]{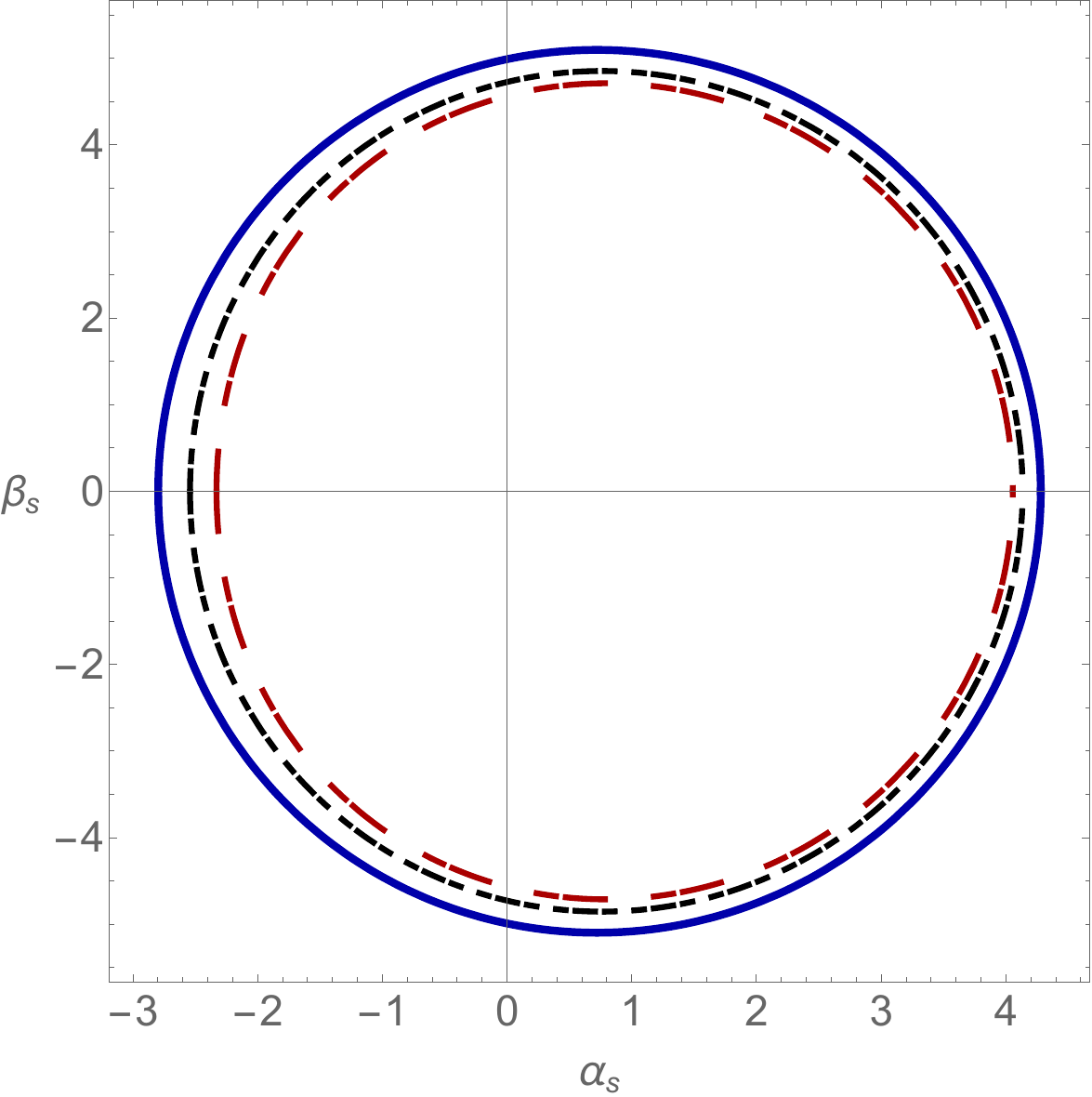}
    \includegraphics[scale = 0.356]{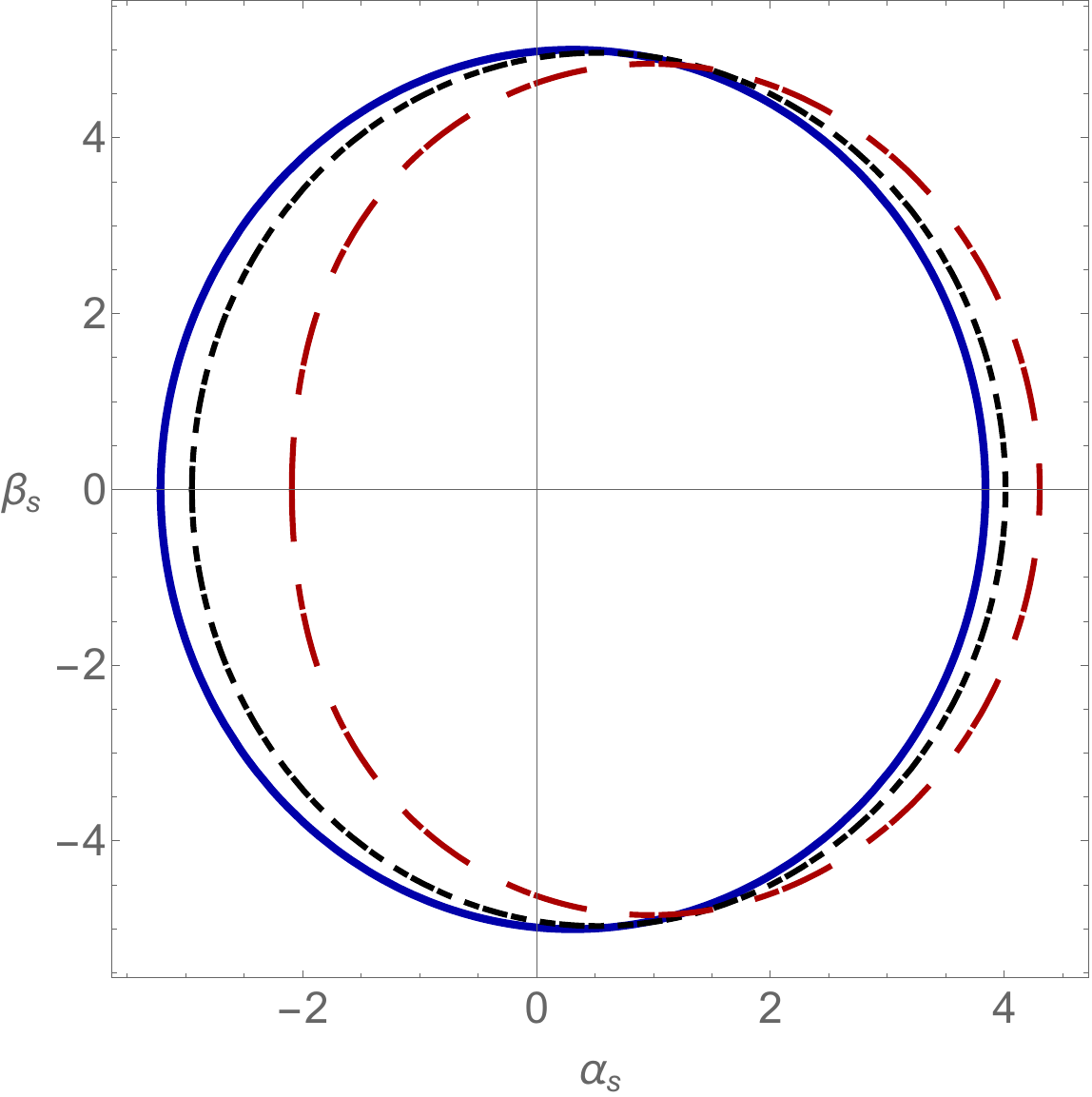}
    \caption{Plot of rotating loop quantum black bounce shadow contours: $\alpha=0.0$ (Kerr - solid blue), $\alpha=0.6$ (dashed black), and $\alpha=0.8$ (long-dashed red), for $a=0.73$ (left panel). For $\alpha=0.5$, we show the variation in the shape of the shadow for different values of $a$, this is, $a=0.3$ (solid blue), $a=0.5$ (dashed black), and $a=0.88$ (long-dashed red) (right panel).}
    \label{shadows}
\end{figure}
If we define the quantities $\xi = J/E$ and $\eta = K/E^2$, and use the fact that unstable circular photon orbits in general rotating spacetime must satisfy $R(r_{\text{ph}}) = 0$, $R'(r_{\text{ph}}) = 0$, and $R''(r_{\text{ph}}) \geq 0$, we obtain (see, for example, \cite{Ghosh2010}):
\begin{equation}
\left[X(r_{\text{ph}}) - a \xi \right]^2 - \Delta(r_{\text{ph}}) \left[\eta + (\xi - a)^2 \right] = 0,
\end{equation}
\begin{equation}
2 X'(r_{\text{ph}}) \left[X(r_{\text{ph}}) - a \xi \right] - \Delta'(r_{\text{ph}}) \left[\eta + (\xi - a)^2 \right] = 0,
\end{equation}
where $r = r_{\text{ph}}$ is the radius of the unstable photon orbit. Furthermore, solving for $\xi$ and $\eta$ from these equations, we find that \cite{Ghosh2010}:
\begin{equation}
\xi = \frac{X_{\text{ph}} \Delta'_{\text{ph}} - 2 \Delta_{\text{ph}} X'_{\text{ph}}}{a \Delta'_{\text{ph}}},
\end{equation}
\begin{equation}
\eta = \frac{4 a^2 X'^2_{\text{ph}} \Delta_{\text{ph}} - \left(X_{\text{ph}} - a^2 \right) \Delta'_{\text{ph}} - 2 X'_{\text{ph}} \Delta_{\text{ph}}^2}{a^2 \Delta'^2_{\text{ph}}},
\end{equation}
where the subscript “ph” indicates evaluation at $r = r_{\text{ph}}$. These equations give the general expressions for the critical impact parameters $\xi$ and $\eta$ of the unstable photon orbits, which describe the shadow's contour.

The unstable photon orbits form the shadow's boundary. The apparent shape of the shadow is obtained using the celestial coordinates $\alpha_s$ and $\beta_s$, which lie in the celestial plane perpendicular to the line joining the observer and the spacetime geometry's center. The coordinates $\alpha_s$ and $\beta_s$ are defined by:
\begin{equation}
\alpha_s = \lim_{r_0 \to \infty} \left(-r_0^2 \sin \theta_0 \frac{d\phi}{dr} \bigg|_{(r_0, \theta_0)} \right),
\end{equation}
\begin{equation}
\beta_s = \lim_{r_0 \to \infty} \left(r_0^2 \frac{d\theta}{dr} \bigg|_{(r_0, \theta_0)} \right),
\end{equation}
where $(r_0, \theta_0)$ are the observer's position coordinates. Taking the limit, we obtain:
\begin{equation}
\alpha_s = -\frac{\xi}{\sin \theta_0},
\end{equation}
\begin{equation}
\beta_s = \pm \sqrt{\eta + a^2 \cos^2 \theta_0 - \xi^2 \cot^2 \theta_0}.
\end{equation}
The shadow is constructed by using the unstable photon orbit radius $r_{\text{ph}}$ as a parameter and plotting $\alpha_s$ and $\beta_s$ using Eqs. (36), (37), (40), and (41).

In Fig. \eqref{shadows}, we depict the shadow of a rotating black bounce for $\theta_0=\pi/2$, which appears smaller as the LQG parameter increases due to quantum corrections in spacetime geometry, for the same rotation parameter, $a$. These corrections cause the photon sphere to shrink, leading to a reduced apparent shadow, consistent with the reduction of the ergospheres. These changes could become relevant discriminators for quantum gravity signatures. 


\section{Conclusion} \label{sec_4}
We studied static and rotating loop quantum black bounces, emphasizing the impact of LQG corrections and the regularization parameter on their properties. Initially, we revisited the spherically symmetric and static black hole exterior solution obtained in \cite{Kelly:2020uwj}, identifying its source as originating from non-linear electrodynamics (NED) with purely electric and magnetic charges. Subsequently, we discussed the energy conditions for these sources, demonstrating that NEC, WEC, and SEC are satisfied, as expected.

We then analytically extended the exterior metric to include the origin, applying the Simpson-Visser prescription to ensure a black hole-wormhole bounce. This results in a spacetime structure significantly different from the original regular black hole solution, yet still dependent on the LQG and bounce parameters. We also analyzed the horizon structure and the existence of singularities. The curvature analysis confirmed the regularity of the spacetime. The examination also revealed that the energy conditions are not satisfied close to the bounce point, but they are met at locations farther from the bounce. However, we demonstrated that increasing the quantum gravity effects through $\alpha$ parameter alleviates this violation. Unlike the LQG black hole, this result suggests that NED alone is insufficient as a source. Indeed, we found that the inclusion of a phantom-type scalar field is necessary. We obtained the $\mathcal{L}(r)^{NED}$ for both electric and magnetic charges. However, we just found $\mathcal{L}(F_{\mu\nu})^{NED}$ for the magnetic case. 

 Additionally, we have explored thermodynamic properties, including the Hawking temperature, entropy, thermal capacity, and Gibbs free energy. Our analysis revealed critical dependencies on the quantum parameter, as well as the possibility of occurring remnants (where $T_H$=0) and phase transitions. Particularly, we analytically found the remnants, which occur when $r_h=\sqrt{4 \alpha ^2-3 r_b^2}/\sqrt{3}$.

We have also derived a more physically interesting rotating loop quantum black bounce solution, analyzing its horizon structure and determining that it remains regular, finding that the transit surface (the throat) of the rotating black-bounce has the spatial geometry of an ellipsoid, which is a natural extension of the spherical surface of the nonrotating case.  This suggests a general result applicable to any stationary solution derived from the static one regularized through the Simpson-Visser prescription. Additionally, our findings indicate that increasing the LQG parameter results in smaller ergospheres and reduced shadows due to quantum corrections in the spacetime geometry. These results have significant implications for observational astrophysics, providing a distinctive signature of quantum gravitational effects that could be detected with future astronomical observations.
\vspace{-0.8cm}
\section*{Acknowledgments}
\hspace{0.5cm} CRM and MSC thank the Conselho Nacional de Desenvolvimento Cient\'{i}fico e Tecnol\'{o}gico (CNPq), Grants no. 308268/2021-6 and 315926/2021-0, respectively.
This work is also supported by the Spanish Grant PID2020-116567GB- C21 funded by MCIN/AEI/10.13039/501100011033 and by the Severo Ochoa Excellence Grant CEX2023-001292-S. 
\newpage
\bibliography{ref.bib}
\end{document}